\documentstyle[prb,aps,multicol]{revtex}
  \renewcommand{\narrowtext}{\begin{multicols}{2} \global\columnwidth20.5pc}
  \renewcommand{\widetext}{\end{multicols} \global\columnwidth42.5pc}
  \newcommand{\wide}{\widetext \noindent \line(200,0){245} \line(0,1){3}\\}

\input{psfig}
\draft{}
\def\be{\begin{equation}}
\def\ee{\end{equation}}

\def\reflh{(\ref{SigmaLowH}) and~(\ref{EpsLowH})}
\def\refapp{(\ref{SigmaLowHSimple}) and~(\ref{EpsLowHSimple})}
\def\shom{\sigma_s^{\rm{hom}}}
\def\sinh{\sigma_s^{\rm{inh}}}
\def\YBCO{YBa$_2$Cu$_3$O$_{7-\delta}$}
\def\NdGaO{NdGaO$_3$}
\def\UD{Ullah and Dorsey}
\def\Tc{$T_c$}
\def\inh{inhomogeneity}
\def\dsp{\delta T_c}
\def\dtr{\Delta T_R}
\def\dtc{\Delta T_c}
\def\dth{\Delta T_{\rm{hom}}}
\def\dte{\Delta T_{\rm{EBIV}}}

\begin{document}

\title
{Temperature and magnetic-field dependence of the conductivity of \YBCO\ films
in the vicinity of superconducting transition: Effect of $T_C$-inhomogeneity  }

\author{D.~V.~Shantsev,\cite{0}
M.~E.~Gaevski, R.~A.~Suris}
\address{Ioffe Physico-Technical Institute, 
Polytechnicheskaya 26, St.Petersburg 194021 Russia}
\author{A.~V.~Bobyl, V.~E.~Gasumyants, O.~L.~Shalaev}
\address{St.Petersburg State Technical University,
Polytechnicheskaya 29, St.Petersburg 195251 Russia}

\maketitle
\begin{abstract}
Temperature and magnetic field dependences of the conductivity of \YBCO\ films
in the transition region are analyzed taking into account spatial inhomogeneity in transition temperature, \Tc.
(i) An expression for the superconducting contribution to
conductivity, $\sigma_s(T,H,T_c)$, of a homogeneous superconductor
for low magnetic fields, $H \ll H_{c2}(0)$, is obtained using the
solution of the Ginzburg-Landau equation in form of perturbation
expansions
[S.Ullah, A.T.Dorsey, Phys. Rev. B {\bf 44}, 262 (1991)].
(ii)~The error in $\sigma_s(T,H,T_c)$ occurring due
to the presence of \Tc-inhomogeneity is calculated
and plotted on an $H-T$ plane diagram.
These calculations use an effective
medium approximation and a Gaussian distribution of \Tc.
(iii)~Measuring the
temperature dependences of a voltage, induced by a focused
electron beam, we determine
spatial distributions of the critical temperature
for \YBCO\ microbridges  with a 2~$\mu$m
resolution.
A typical \Tc-distribution dispersion is found
to be $\approx$1~K.
For such dispersion, error in $\sigma_s(T,H,T_c)$
due to \Tc-inhomogeneity exceeds 30\% for magnetic fields
$H<1$~T and temperatures $|T-T_c|<0.5$~K .
(iv)~Experimental $R(T,H)$ dependences of resistance are well described by
a numerical solution of a
set of Kirchoff equations for the resistor network
based on the measured spatial distributions
of \Tc\ and the expression for $\sigma_s(T,H,T_c)$.
\end{abstract}


\narrowtext
\section{Introduction}

The complicated crystal structure of high-$T_c$
superconductors (HTSC) leads to their substantial spatial
inhomogeneity, which is especially important because of the
very short coherence length, $\xi$.
Inhomogeneities with spatial scale much larger than $\xi$
allow for an inhomogeneous distribution
of the critical temperature, \Tc, which affects
properties of HTSC
in the superconducting transition vicinity.
As a result,
it is often difficult to establish whether observed
behavior of superconductors arise from intrinsic
properties or from spatial \inh. This impedes
analysis of experimental data in the transition region, which is
often used to determine  microscopic
superconducting parameters and the
mechanism of superconductivity.

The most obvious origin of \Tc-\inh\ is variation in
oxygen content over the sample.
For YBa$_2$Cu$_3$O$_{7-\delta}$ (YBCO),
$T_c$ is a relatively weak function of $\delta$
at $6.85 < 7-\delta < 7$ (so called 90~K plateau)
and falls abruptly at higher
$\delta$.\cite{PhysCSiddique,rusGasumants}
Even $\delta$ variations within the 90~K plateau
can lead to $\approx$ 1~K variations in \Tc.
Meanwhile, experimental x-ray data show that $\delta$ variation
can be substantially higher even for crystals exhibiting
excellent transport properties.\cite{Browning}
Another origin of \Tc-inhomogeneity  is variation in cation (Y, Ba, Cu)
composition. This origin can be dominant in
thin YBCO films, as shown by
simultaneous spatially resolved studies of
cation composition and \Tc\ using electron probe microanalysis and
low-temperature scanning electron microscopy
(LTSEM), respectively.\cite{cations,3methods}
Elastic stresses around structural defects
can also lead to \Tc-inhomogeneity  due to a
strong pressure dependence of \Tc\ in HTSC
compounds.\cite{pressure}  \Tc\ increase due to edge
dislocations and low-angle grain boundaries was
calculated to be $\sim 1$~K.\cite{Gurevich}
Spatial variations of the $c$-axis lattice parameter
revealed by x-ray studies in YBCO films with
almost uniform oxygen content
also suggest stress-induced \Tc-inhomogeneity.\cite{JAP}

Presence of \Tc-inhomogeneity  manifests itself in various HTSC properties.
Temperature dependence of the depinning current density
in YBCO crystals implies that
pinning sites are induced by  spatial variations
of \Tc.\cite{pinningWen} Systematic studies of
YBCO crystal  magnetization curves suggest
the presence of local regions with reduced
oxygen content, leading to the so called
peak effect.\cite{Kupfer,Zhukov,Osofsky}
Meanwhile, \Tc-inhomogeneity  should have even greater impact on
temperature dependences of transport coefficients
just above the superconducting transition.
This is confirmed by experimental data on conductivity,
magnetoconductivity, and the Hall coefficient at
temperatures $T \ge T_c+2$~K from
Refs.\onlinecite{PomVidal,Lang,Lang95}
which were explained by
assuming a Gaussian distribution of \Tc\ with dispersion
in the range 0.6-2.3~K.
However, these results can only serve
as an {\em indirect} indication of the presence of
\Tc-inhomogeneity, due to a lack of experimental data about
real \Tc-distribution in the samples.
Moreover, the temperature region in the
vicinity of the superconducting transition,
where \Tc-inhomogeneity  is especially
important, was not considered.

A step forward has been made in Ref.~\onlinecite{MosSST} where 
resistor network calculations are used to analyse
current density redistributions in
\Tc-inhomogeneous supercondictor in the transition region.
It is shown that some anomalities in the 
temperature dependence of
in-plane magnetoresistivity, such as negative
magnetoresistivity excess, which are usually
attributed to intrinsic effects can be quantitatively
expalined by nonuniform \Tc-distribution.

In the present work, we investigate the influence of \Tc-inhomogeneity
on properties of HTSC throughout the transition region and
analyze  experimentally determined spatial
distributions of \Tc\ (\Tc-maps).
Measuring \Tc-maps of YBCO films by LTSEM with 2~$\mu$m
resolution, we reveal $\approx$ 1~K scatter of \Tc\ over
the films. To calculate the effective conductivity of such an
inhomogeneous material, one needs the expression for
conductivity $\sigma(T,H,T_c)$ of a uniform
superconductor valid throughout the transition region.
Such an expression was obtained in Ref.~\onlinecite{UD}
by solving the time-dependent Ginzburg-Landau equation with
Lawrence and Doniach Hamiltonian in the Hartree
approximation.

It is well-known that magnetic field leads to a broadening
of the superconducting transition which is roughly proportional
to $\left(dH_{c2}/dT \right)^{-1}_{T=T_c}\approx 0.5$~K/T.\cite{Welp} For
fields $H>2$~T, this broadening dominates over
inhomogeneous broadening due to scatter of local values of \Tc.
Therefore, we are interested in low
fields, $H \le 2$~T, where the influence of \Tc-inhomogeneity  is essential.
Moreover, this range of magnetic fields is actual
for most HTSC applications.
Unfortunately, the final formula for conductivity obtained in
Ref.~\onlinecite{UD} is only valid for high magnetic fields.
In the present work, we deduce expressions for conductivity valid for
low magnetic fields, $H \ll H_{c2}$, from the perturbation expansions
in Ref.~\onlinecite{UD}.

The present paper is organized as follows. In Sec.~II the
expression for the Cooper pair conductivity,
$\sigma_s(T,H,T_c)$, of a homogeneous
superconductor is derived. In Sec.~III we discuss
methods to
calculate the effective conductivity of an inhomogeneous
superconductor. In Sec.~IV
the error in the value of $\sigma_s(T,H,T_c)$ occurring due
to the presence of \Tc-inhomogeneity is calculated and the results
are plotted on the $H-T$ plane diagram.
Sec.~V describes the samples and
experimental techniques.
In Sec.~VI,  measured spatial distributions of \Tc\ are
analyzed and the broadening of the
superconducting transition due to \Tc-inhomogeneity  is discussed.
Finally, the experimental $R(T,H)$ dependences of resistance are
interpreted on the basis of measured \Tc-distributions and
the expression for $\sigma_s(T,H,T_c)$ derived in Sec.~II.

\section{Conductivity of a homogeneous superconductor}

To describe the temperature dependence of conductivity of a homogeneous superconductor throughout the transition region we 
employ the results obtained by Ullah and Dorsey.\cite{UD} They studied the time-dependent Ginzburg-Landau equation for 
anisotropic superconductor with the Hamiltonian introduced by Lawrence and Doniach\cite{LD} and an additional noise term. The 
magnetic field $H$ was assumed to be applied along the $c$-axis. Using the Hartree approximation Ullah and Dorsey obtained  
expressions for the transport coefficients which gave smooth interpolation between the high-temperature regime dominated by  
Gaussian fluctuations and low-temperature  flux-flow regime. The expression for the Cooper pair conductivity in linear order to 
electric field was obtained in the form of two coupled perturbation expansions\cite{UD}:

\be
\sigma_s=\sigma_0\sum_{n=0}^N(n+1)(A_n+A_{n+1}-2A_{n+1/2}),
\label{SigmaRyad}
\ee
where\cite{UDerror}
\be
A_n = A_n(\tilde \epsilon_H,h)=
\left[
(\tilde \epsilon_H+2hn)(1+d^2(\tilde \epsilon_H+2hn))
\right]^{-1/2},
\label{An}
\ee
and
\be
\epsilon_H=\tilde \epsilon_H-\Omega T h\sum_{n=0}^N A_n.
\label{EpsRyad}
\ee
Here $h=H/H_{c2}(0)$, and $\epsilon_H$ is a field-dependent
dimensionless temperature, $\epsilon_H=T/T_c-1+h$. Further,
\be
 \Omega= \frac{8\pi^2 \left( 2\kappa^2 -1 \right)
\xi_c(0) k_B}  {\gamma^2 \phi_0^2},
\label{Omega}
\ee
$\xi_{ab}(0)$ and $\xi_c(0)$ are the correlation lengths
in
the CuO plane and transverse to it:
$\xi_{ab}(0)=\hbar/\sqrt{2m_{ab}\alpha_0}$
(with similar relation for $\xi_c(0)$);
 $m_{ab}$ is the Cooper pair mass in CuO plane,
$\alpha_0$ is related to
the parameter $\alpha$ in Ginzburg-Landau Hamiltonian as
$\alpha= \alpha_0 \left( T/T_c-1 \right)$,
$d=s/2\xi_c(0)$, where $s$ denotes spacing between CuO planes,
$\gamma=\xi_c(0)/\xi_{ab}(0)$ is
an anisotropy parameter,
$\phi_0$ is the flux quantum,
$\kappa$ is the Ginzburg-Landau parameter,
$H_{c2}(0)=\phi_0 / 2\pi \xi_{ab}^2(0)$, $N = 1/h$,
and $\sigma_0$ is a constant
with dimensionality of a conductivity.

In order to avoid summation in the above expressions,
the following approximation suggested in Ref.~\onlinecite{UD}
is usually used.
For high magnetic field ($\tilde \epsilon_H \ll 2h$) and
3D case ($d^2\tilde \epsilon_H \ll 1$)
only terms containing $A_0$ are left in
Eqs.~(\ref{SigmaRyad}) and (\ref{EpsRyad}) yielding
$\sigma_s = \sigma_0 / \sqrt{\tilde \epsilon_H}$
and
$\epsilon_H = \tilde \epsilon_H - \Omega h / \sqrt{\tilde \epsilon_H}$
respectively.
This leads to
\begin{equation}
 \sigma_s = \frac{\sigma_0 }{(\Omega T h)^{1/3}}\,\, {\cal F}
 \left[  \frac{\epsilon_H}{\left( \Omega T h \right)^{2/3}}
\right],
\label{Ullah}
\end{equation}
where function ${\cal F}(x)$ satisfies cubic equation,
\begin{equation}
 x{\cal F}^2=1-{\cal F}^3 .
\label{cubic}
\end{equation}
The solution of this equation can be written down as
\begin{eqnarray}
{\cal F}(x)&=&\theta +x^2/(9 \theta)-x/3, \nonumber \\ 
\theta &=&\left( 1/2-x^3/27+\sqrt{(27-4x^3)/108}
\right)^{1/3}.
\label{F3D}
\end{eqnarray}
The function ${\cal F}(x)$ is equivalent to the function $F_{3D}$
in Ref.~\onlinecite{UD} though
expressions~(\ref{cubic}) and~(\ref{F3D})
for this function are not presented there.

Despite Eq.~(\ref{Ullah}) is widely used, its applicability
range needs a special discussion.
Indeed, since the series (\ref{EpsRyad}) are diverging,
omitting all terms except the first one
is hardly permissible.
At least, it is obviously incorrect
if the condition $\tilde \epsilon_H \ll 2h$~ does not hold.
Hence, for magnetic fields we deal with, $H < 2$~T,
the high-field approximation of Ref.~\onlinecite{UD}
is not valid except for the low-temperature part of
the superconducting transition.
As argued in Ref.~\onlinecite{UD},
Eq.~\ref{Ullah} can also be considered as a scaling relation with unknown scaling function ${\cal F}$ and then it is valid in a 
wider range of magnetic fields.
However, simple numerical calculations show that, e. g.,
for \YBCO, the scaling does not work for fields $H<2$~T.
Therefore, we derive new, low-field approximation
for conductivity from Eqs.~(\ref{SigmaRyad}) and
(\ref{EpsRyad}).
For $h \ll 1$
one can replace summation for $n\ge 1$ by integration 
using Euler-Maklaurin formula\cite{EM} and
obtain the following analytical expressions for
conductivity:

\begin{eqnarray}
\sigma_s / \sigma_0&=&
A_0 - 2 A_{1/2} + \frac94  A_1 - 2 A_{3/2} + A_2
\nonumber \\&&
- {3+8d^2 \over 4(2+4d^2)^{3/2}}
+ {3h\over 8}  (1+2d^2(\tilde \epsilon_H + 2h)) A_1^3,
\label{SigmaLowH}
\end{eqnarray}
\begin{eqnarray}
\epsilon_H&=&\tilde \epsilon_H-\frac{\Omega T} {2}
\left(2h A_0 +
h A_1+
{h\over\sqrt{2(1+2d^2)}}
\right. \nonumber \\&&\left.
+{1\over d} \ln\left({d+4d^3+2d^2\sqrt{2+4d^2}
\over 2d^3\tilde \epsilon_H+d+2d^2 / A_1}\right)\right)
\label{EpsLowH}
\end{eqnarray}

Equations \reflh\ present non-explicit dependence of
conductivity for a homogeneous superconductor
on temperature and magnetic field.
They are derived without any assumptions
about 3D or 2D character of superconductivity
and therefore applicable for
arbitrary anisotropy parameter.
These equations are used below for calculations presented
in Sec.~III and~V.

Under conditions
\be
 \tilde \epsilon_H,h \ll 1/d^2  \ll 1
\label{d}
\ee
equations \reflh\ can be substantially simplified yielding
an {\em explicit} expression for the Cooper pair conductivity:
\begin{eqnarray}
\frac{\sigma_s}{ \sigma_0}=
{1\over\sqrt{\tilde \epsilon_H}}- {2\over\sqrt{\tilde \epsilon_H+h}}+
{1\over\sqrt{\tilde \epsilon_H+2 h}}
+{7 h + 2\tilde \epsilon_H \over 8
 (\tilde \epsilon_H+2 h)^{3/2}}
\label{SigmaLowHSimple}
\end{eqnarray}
\begin{eqnarray}
\tilde \epsilon_H &=& (\Omega T h)^{2/3}\,\, {\cal F}^{-2}
 \left[ \frac {(1 + \Lambda)\ T / T_c  -1 + h }{ (\Omega T h)^{2/3}  }
 \right], 
\nonumber \\
 \Lambda &=&  \Omega T_c \ \frac{\ln 8d^2}{2d},
\label{EpsLowHSimple}
\end{eqnarray}
where ${\cal F}$ is the function defined by Eqs.~(\ref{cubic})
or~(\ref{F3D}).
Another advantage of these equations is that 
the conductivity depends on $d$ only
through parameter $\Lambda$ which characterizes the shift
of the apparent transition temperature
with respect to $T_c$. This shift is always present in the Hartree approximation as a result of renormalization of the
parameter $\alpha$ in the Ginzburg-Landau Hamiltonian.

Eqs.\reflh\ as well as Eqs.\refapp\
defining temperature and magnetic field dependence of
conductivity  contain only two key parameters:
$d$ and $\Omega$.
Parameter $d$ depends on the value of
coherence length $\xi_c(0)$ which is not known well.
An accurate determination of $\xi_c(0)$ is difficult
because of bilayered structure of the YBCO unit cell.
A systematic analysis performed in Ref.~\onlinecite{bilayer}
shows that a good approximation for YBCO is the assumption
of equally spaced CuO layers with interlayer distance $s=6\ \AA$.
Then, a simultaneous fitting of conductivity, magnetoconductivity and
susceptibility data gives
$\xi_c(0)=1.2\AA$ leading to $d=2.5$.\cite{bilayer}
This value does not contradict to results of
other works\cite{rice,xi,MTargue} assuming the distance between
superconducting layers $s\approx 11.7\ \AA$. They give
values of $\xi_c(0)$ in the range between 1.3 and 3~$\AA$
which corresponds to $1.9<d<4.5$.
Obviously, this uncertainty in $d$ is very large.
Fortunately, it is not so important in the
range of $H$ and $T$ that we consider.
Indeed, for YBCO the conditions (\ref{d}) are satisfied
within several kelvins around the
transition unless the magnetic field is very high.
Then, as follows from Eqs.~\refapp,
\begin{figure}[p]
\centerline{ \psfig{figure=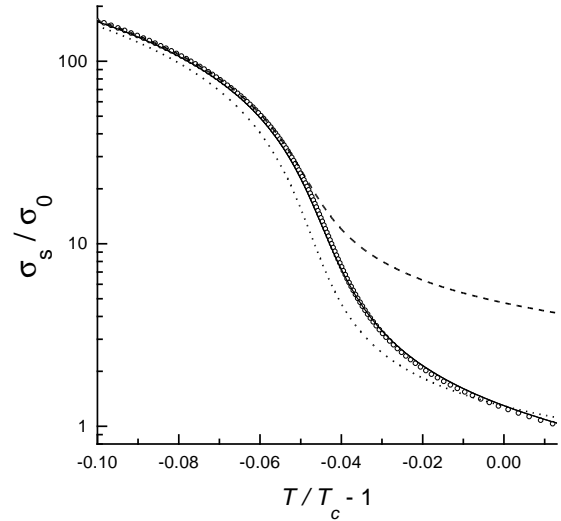,width=8cm}}
\caption{\label{f_SumApp}
Temperature dependence of Cooper pair
conductivity for YBCO in magnetic field
$H = 2$~T. Symbols show the exact
result of UD model,\protect\cite{UD}
Eqs.~(\protect\ref{SigmaRyad}), and~(\protect\ref{EpsRyad}); dashed line is
the high-field approximation proposed by \UD,
Eqs.~(\protect\ref{Ullah}) and~(\protect\ref{F3D})
(shifted along $x$-axis by $-\Lambda$); solid line is
the low-field approximation proposed in the present paper,
Eqs.~\protect\reflh; dotted line is
an approximate explicit expression for low
fields given by Eqs.~\protect\refapp.}
\end{figure}

\noindent
change in $d$ leads only to a shift of apparent transition 
temperature but does not affect the shape of the
transition curve.
The second parameter
$\Omega$ depends on the coherence length $\xi_{ab}(0)$ and the
Ginzburg-Landau parameter $\kappa$.
The former was also taken from Ref.~\onlinecite{bilayer},
$\xi_{ab}=11\ \AA$. Meanwhile, it is quite difficult to find in literature
an accurate estimate for $\kappa$. We used the value $\kappa=30$ providing
the best fit to our experimental data for YBCO films.
Then, from equation (\ref{Omega}) one obtains $\Omega = 5\times 10^{-4}$.

Figure~\ref{f_SumApp} presents comparison of the temperature
dependences of Cooper pair conductivity
given by straightforward summation in (\ref{SigmaRyad})
and~(\ref{EpsRyad}),
and by two analytical approximations:
high-field approximation, Eqs.~(\ref{Ullah}) and
(\ref{F3D}),
proposed by \UD\ and
low-field approximation, Eqs.~(\ref{SigmaLowH}) and
(\ref{EpsLowH}),
suggested in the present work. An
approximate explicit expression for low
fields given by Eqs.~\refapp\ is also shown as dotted line.
For $H=2$~T which is the case shown in the figure,
the low-field approximation
is far more accurate than the high-field one.
For lower magnetic fields the deviation between
the result of exact summation and the low-field approximation
is almost indistinguishable.
By contrast, the high-field approximation fails
for temperatures $T \gg T_c$ where
it four times overestimates the result of exact summation which is
$\sigma = \sigma_0/(4\sqrt{\epsilon_H})$.

Figure clearly shows that 
the apparent transition temperature
is shifted downward from $T_c$.
For given set of parameters the dimensionless
shift is $\Lambda \approx 0.03$.
In order to avoid confusion,  the data in all figures below are
shifted along $x$-axis so that \Tc\
corresponds to the apparent transition temperature.
It should be also noted that the shift $\Lambda$ 
does not enter the 
high-field approximation suggested by Ullah and Dorsey, 
Eqs.~(\ref{Ullah}) and (\ref{F3D}).
This approximation predicts transition at $T=T_c$
in contradiction to basic equations of UD model, 
Eqs.~(\ref{SigmaRyad}) and~(\ref{EpsRyad}).
In order to make the high-field approximation merge
all other curves in Fig.~\ref{f_SumApp} at least at low $T$, 
we had to shift the corresponding 
dashed curve on the value $\Lambda$ ``by hand".

The constant $\sigma_0$ entering UD model
depends on a phenomenological quantity,
the relaxation rate of the order parameter.
It is natural to estimate $\sigma_0$ using
well-known Aslamazov-Larkin result\cite{AL} for
high-temperature asymptotic in 3D case:
$ \sigma_s^{3D} = {e^2}/{32\hbar\xi(0)\sqrt{\epsilon_H}}$.
Thus, we have
\begin{equation}
 \sigma_0 = e^2 / \, 8\hbar\xi(0)
\label{sigma0}
\end{equation}

Let us now discuss the applicability range for
the results obtained in this Section.
The indirect (Maki-Thompson) contribution to
the order parameter fluctuations \cite{MT}
is not taken into account in the UD model.
However, there are grounds to believe that
neglecting Maki-Thompson term would not affect the
results obtained in  the transition region
(few Kelvins around \Tc) since the
direct Aslamazov-Larkin process is dominant over
the indirect one in this temperature range.\cite{MTargue}
One should also keep in mind that the UD model does not
take into account vortex pinning and predicts flux-flow behavior
in the limit of low temperatures.
Therefore, it cannot be used at temperatures well below $T_c$,
where the current-voltage characteristics are nonlinear.

\section{Account of \Tc-inhomogeneity}

Let us now consider how the properties of a
superconductor can be affected by spatially inhomogeneous
distribution of critical temperature. First, we suppose that the
correlation length $r_c$ of \Tc-distribution is so large that
the temperature region near \Tc\ where $\xi(T)>r_c$ can be
ignored. This assumption seems to be quite reasonable
since the coherence length of HTSC is much smaller than
$r_c$ obtained from LTSEM data, see Tab.~1.
The condition $r_c \gg \xi$ makes it possible
to ignore the correlation between the superconducting order
parameter in adjacent fragments and to consider them
independently. Therefore, the expression for the conductivity
obtained in the previous section for a homogeneous superconductor
can be used to describe local conductivity $\sigma(T,H,T_c)$
of a homogeneous fragment with given \Tc.

The straightforward way to determine the conductivity
$\sigma^{\rm inh}(T,H)$ of a \Tc-inhomogeneous superconductor is
to start from the spatial distribution of \Tc\ over the
sample. The value of $\sigma^{\rm inh}(T,H)$ can be
determined exactly from the values of local
conductivities $\sigma(T,H,T_c)$. In this work we
determined the spatial distributions of the critical
temperature in YBCO films using LTSEM
(see Sec.~V).
This method, however, leads to
the lack of information about small-scale
inhomogeneities with $r_c \le r_{\exp}$, where $r_{\exp}$ is
the spatial resolution of the technique.
Therefore, if small-scale inhomogeneity is essential,
or if spatial distribution of \Tc\ is unknown,
a Gaussian \Tc-distribution function together with, e. g.,
effective medium approach can be used to find $\sigma^{\rm inh}(T,H)$.

The problem of conductivity of an inhomogeneous medium
has the exact analytical solution only for a special case of symmetric
distribution of phases in 2D system.\cite{Dyhne} In the
general case one has to use some approximation. According
to the effective medium approach\cite{EMA} (EMA), the
conductivity $\sigma^{\rm inh}(T,H)$ is given by the
solution of the equation

\begin{equation}
 \int \frac{ \sigma^{\rm inh}(T,H) - \sigma(T,H,T_c) }
 { (D-1)\, \sigma^{\rm inh}(T,H) + \sigma(T,H,T_c) } f(T_c)
dT_c = 0,
\label{EMA}
\end{equation}
where $D$ is the dimensionality of the system.  Here
$f(T_c)$ is a distribution function of critical temperature  over
the sample which shows the relative volume occupied
by fragments with given \Tc. Despite the apparent
simplicity, EMA gives rather high accuracy
(up to few percents) unless
the system is in the very vicinity of the percolation
threshold.\cite{kir}  In the case of thin film samples
with thickness less than the correlation length of \Tc-inhomogeneity,
$r_c$, EMA expression (\ref{EMA}) with dimensionality
$D$=2 should be used.  It should be emphasized that this
dimensionality has nothing to do with the dimensionality
of the superconducting properties mentioned in relation
with formula (\ref{Ullah}); the first one depends on the
geometry of the sample,
while the latter is associated with anisotropy of the
crystal structure.

\section{$H-T$ diagrams}

In this section we estimate the effect of \Tc-inhomogeneity
on the apparent value of the Cooper pair
conductivity in the vicinity of the
superconducting transition.
Usually, experimental data on $\sigma(T,H)$-dependences
in the transition region are studied first by subtracting
the conductivity of normal electrons, $\sigma_n$, and then
analyzing the remaining conductivity of Cooper pairs,
$\sigma_s$.
In the case of inhomogeneous sample such procedure would
lead
to an error in $\sigma_s$: its apparent value determined
from experimental data would be different from that for a
homogeneous superconductor.
To quantitatively estimate this error we consider two
samples: uniform, with critical temperature $T_{c0}$, and
inhomogeneous one with a Gaussian distribution of
critical temperatures with average $T_{c0}$ and
dispersion $\dsp$:
\be
 f(T_c) ={1\over\sqrt{2\pi}\,\dsp} \exp \left( - \frac{
\left(T_c - T_{c0} \right)^2 }
 {2\,\dsp ^2} \right).
\label{gauss}
\ee
Now two quantities, $\shom(T,H)$ and $\sinh(T,H)$ can be
compared. $\shom$ is the Cooper pair
conductivity for homogeneous sample given by the
expressions~\reflh\ obtained on the basis of UD model.
The conductivity $\sigma^{\rm{inh}}$ of the inhomogeneous sample is
determined by EMA formula~(\ref{EMA}) with $f(T_c)$ being
Gaussian distribution function~(\ref{gauss}), and local
conductivities defined as sum of the superconducting, $\shom$,
and normal, $\sigma_n$, contributions. Then, one should
subtract the normal contribution  from $\sigma^{\rm{inh}}$ and obtain
the apparent superconducting contribution to conductivity for the
inhomogeneous sample:
\be
\sinh(T,H) = \sigma^{\rm{inh}}(T,H) - \sigma_n(T,H).
\label{sinh}
\ee

Further, to proceed with calculations some assumptions
are needed about the temperature and magnetic field
dependences of $\sigma_n$. We neglect the
magnetoresistance of HTSC in the normal state which is
very small and use a linear approximation for the
temperature dependence of the resistivity:
\be
 \sigma_n(T,H) = \sigma_n(T,0) = 1 / \left( C_1+C_2 T \right).
\label{sigman}
\ee

\wide
\begin{figure}[p]
\centerline{ \psfig{figure=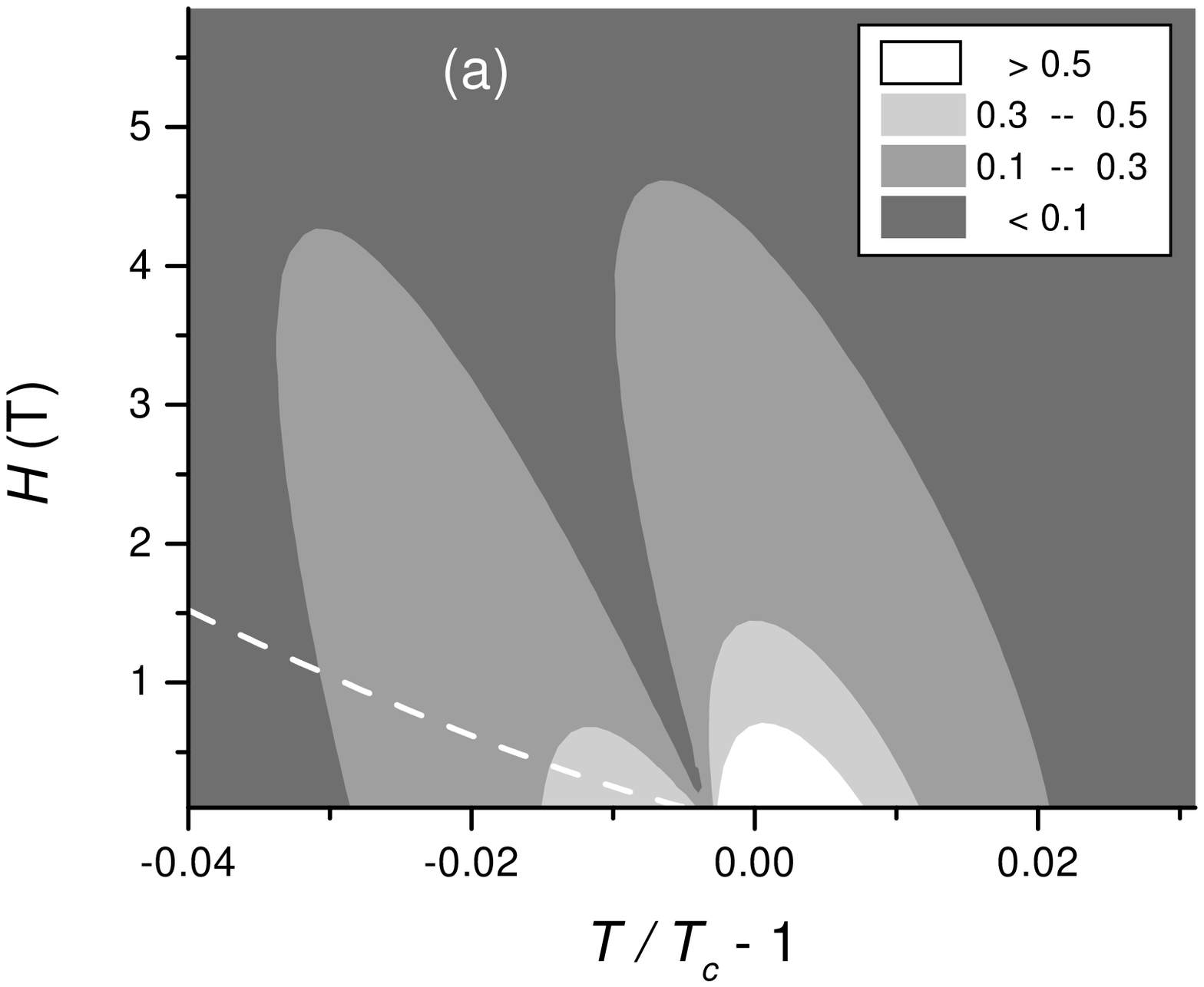,width=10.5cm}
\psfig{figure=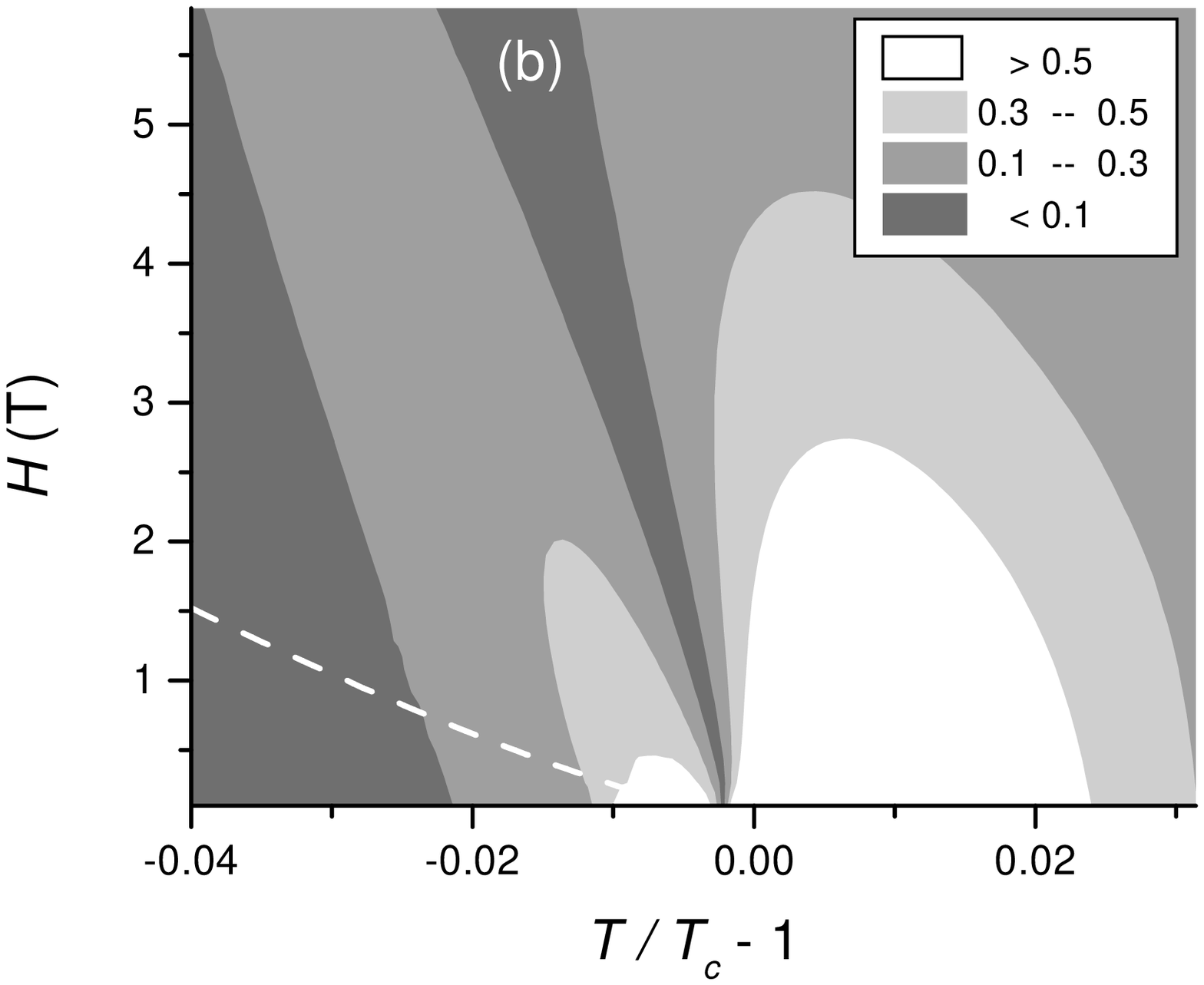,width=10.5cm}}
\caption{
Diagrams in the ``magnetic field''---``temperature'' plane
illustrating the
effect of \Tc-inhomogeneity  on the apparent value of Cooper pair
conductivity, $\sigma_s$.
Absolute values of the reduced difference of (a)
conductivities $|\sinh/\shom-1|$ and (b)
magnetoconductivities $|\left(\sinh \right)_H' / \left(\shom
\right)_H'- 1|$ for homogeneous and inhomogeneous
superconductors are shown.
Calculations are based on Eqs.~\protect\reflh.
For inhomogeneous superconductor, the calculations use
an effective medium approach, Eq.~(\protect\ref{EMA}), and
a Gaussian distribution of \Tc\ with
dispersion $\dsp=1$~K.
Brighter regions correspond to a stronger influence of
\Tc-inhomogeneity.
The influence is maximal in low magnetic fields
and in the vicinity of \Tc.
Presented results are not valid
below the white dashed lines which correspond to the melting transition
of vortex lattice (the data are taken from
Refs.~\protect\onlinecite{meltingPRB,meltingPRL}).}
\label{f_fi2d}
\end{figure}
\narrowtext

The key parameter for calculations is the
dispersion of Gaussian \Tc-distribution (\ref{gauss}), $\dsp$.
We used the value $\dsp=1$~K which
is approximately the average dispersion
for studied YBCO films
determined from their \Tc-maps. The values for
$\Omega$ and $d$ were the same as in Sec.~II, other
parameters were: $C_1$=1.06~$\mu\Omega$m,
$C_2$=0.0035~$\mu\Omega$m/K (see Tab.~1),
$T_c$=90~K, $D$=2.
We also assume that the inhomogeneity of a
superconductor manifests itself in inhomogeneity of
the critical temperature only, while all the other
superconducting parameters and the normal state
conductivity are supposed to be uniform.

It is convenient to consider $H$--$T$ diagram which shows
the absolute value of the relative difference: $|\sinh /
\shom - 1|$, see Fig.~\ref{f_fi2d}a. The effect of \Tc-inhomogeneity
on the magnetoconductivity is illustrated by Fig.~\ref{f_fi2d}b
showing the same diagram for the quantity $|\left(\sinh
\right)_H' / \left(\shom \right)_H'- 1|$, where $(\sigma_s)'_H$ denotes
partial derivative of conductivity with respect
to magnetic field. Brighter regions on the
diagrams correspond to larger values, i.e., to
stronger influence of \Tc-inhomogeneity  on the
values of $\sigma_s$ and $(\sigma_s)_H'$. The influence
becomes crucial in
a $1$~K--wide region around \Tc\ and for magnetic fields
$H < 0.5$~T where ignoring \Tc-inhomogeneity  would lead to $\sim 50\%$
error in $\sigma_s$.
The following
conclusions can be drawn from the diagrams:

(i) \Tc-inhomogeneity plays greater role in the very
vicinity of the transition; far from the transition the
difference in local \Tc's is small compared to $|T-T_c|$
and, hence, not so important.

(ii) \Tc-inhomogeneity plays greater role in low
magnetic fields. The application of magnetic field leads
to a broadening of the transition even in a homogeneous
superconductor.
Since for most HTSC~ $dH_{c2}/dT \approx -2$~T/K\
at~ $T=T_c$,\cite{Welp} one can roughly estimate the increase in the
transition width as one degree for $H$ increase of 2~T.
Therefore, for fields $H>2$~T the dispersion in
critical temperatures $\dsp \approx 1$~K is masked by $H$-induced
broadening of the transition.

(iii) \Tc-inhomogeneity  has greater effect on the
magnetoconductivity of a
superconductor than on its conductivity. {}From
practical point of view it is often preferable to
analyze experimental data on magnetoconductivity
rather than on conductivity.
This is because the contribution of normal electrons to
magnetoconductivity is negligible in the vicinity of \Tc, while
the analysis of conductivity data always requires account of the
normal conductivity and, hence, additional assumptions about its
temperature  dependence. However, as follows from the diagrams,
the analysis of magnetoconductivity data needs more careful
account of \Tc-inhomogeneity. The reason for that, as was earlier
noted by Lang et al.,\cite{Lang95} lies in the stronger
dependence of magnetoconductivity on \Tc, e.g.,
for high temperatures, $T\gg T_c$,
one has $\sigma_s \propto \left( T-T_c \right)^{-1/2}$,
while $\partial\sigma_s/\partial H \propto \left( T-T_c
\right)^{-3/2}$.

The dashed line in Fig.~\ref{f_fi2d} corresponds to
the melting transition of the Abrikosov vortex lattice
as determined from experiments on
YBCO crystals.~\cite{meltingPRB,meltingPRL}
It is remarkable that different methods,
neutron small angle scattering,\cite{meltingPRB}
as well as magnetization and transport measurements,\cite{meltingPRL}
yield the same position of the melting line.
We believe that it can serve as a rough estimate of
the applicability range of
the UD model. Below this line, our results
obtained on the basis of the UD model are not valid.

\section{Samples and Experimental details}

\YBCO\ films with thickness of 0.2~$\mu$m were grown by
dc magnetron sputtering on NdGaO$_3$, AlLaO$_3$ and MgO
substrata. The details of the procedure are described
elsewhere.\cite{JAP} X-ray data have shown the presence
of only (00l) reflexes
confirming $c$-orientation of the films. The Raman
spectroscopy analysis has revealed their high
epitaxiality. Microbridges of 500$\times$50~$\mu$m size
were formed by a standard photolithography. Six
samples were investigated; some important parameters are
presented in Tab.~1.

The temperature dependences of the resistivity were
measured at driving current 1~mA and magnetic fields
$H$=0, 0.3, 0.6, 0.9~T~ applied along the $c$-axis.
Measurements were done inside a temperature stabilized
Oxford He flow cryostat (model CF-1200) under helium
atmosphere, using the standard four-probe dc method, a
Keithly  220 programmable current source and a Keithly
182 sensitive digital voltmeter. Contacts to the samples
were made by thin gold wires attached to the sample
surface by silver paste. The temperature inside the
cryostat was controlled and stabilized by an Oxford
programmable temperature controller ITC4 with accuracy up
to 0.01~K. The temperature of the sample was measured by
copper-constantan thermocouple; voltage was read by a DMM5000
integrating digital multimeter. The measurement was
started when the sample was in the normal state (at least
40~K above \Tc) and performed during a slow cooling
procedure down to zero resistivity of the sample. Then
the sample was heated and the measurement was repeated at
another value of magnetic field. The accuracy of the
voltage measurements was about 10~nV.

The LTSEM measurements were carried out with an automated scanning
electron microscope CamScan Series 4-88 DV100.
The microscope is equipped with a cooling sample stage,
its temperature can be lowered down to 77~K
using an Oxford N flow cryostat.
The temperature is maintained in the range 77-350~K 
With accuracy up to
0.1~K by a temperature controller ITC4.
The bias current was varied from 0.2
to 2.0~mA so that its value was large enough to detect
electron beam induced voltage (EBIV) and
small enough to avoid distortion of the superconducting
transition. EBIV was measured using the standard four-probe method.
A precision instrumentation
amplifier incorporated into the microscope chamber was used to increase
the signal-to-noise ratio.
To extract the local EBIV signal, lock-in detection was used with a
beam-modulation frequency of 1 kHz. The electron beam current was
10$^{-8}$~A, while the acceleration voltage was 10~kV.

\begin{figure}[p]
\centerline{ \psfig{figure=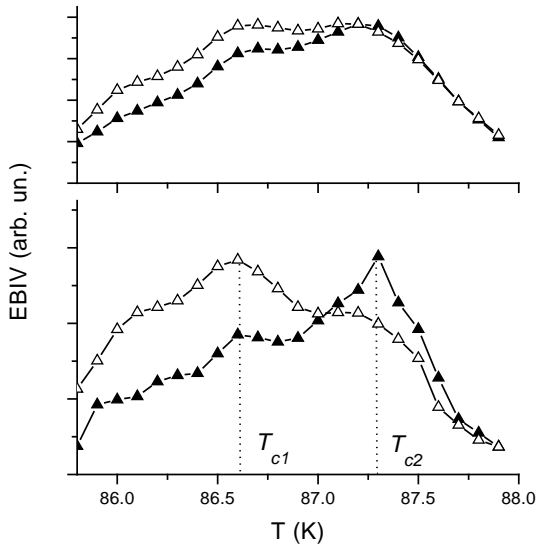,width=8cm}}
\caption{\label{f_ebiv}
Temperature dependences of electron beam induced voltage
measured by LTSEM for two regions of sample~1
separated by 5~$\mu$m distance. The upper panel shows raw
signals, the lower panel shows the same dependences after
deconvolution procedure. Local values of the critical temperature,
$T_{c1}$ and $T_{c2}$, are determined by
positions of the peaks. }
\end{figure}

\begin{figure}[p]
\centerline{ \psfig{figure=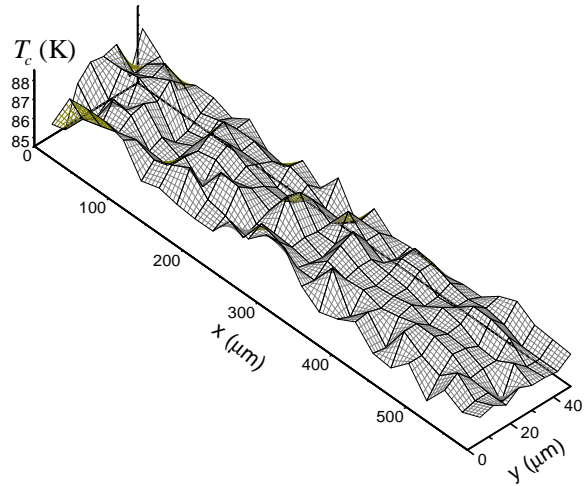,width=8cm}}
\caption{\label{f_tcmap}
Spatial distribution of critical temperature in sample~1
determined from LTSEM data. Distribution is
smoothed with respect to the initial one measured with
2~$\mu$m resolution.}
\end{figure}

\begin{figure}[p]
\centerline{ \psfig{figure=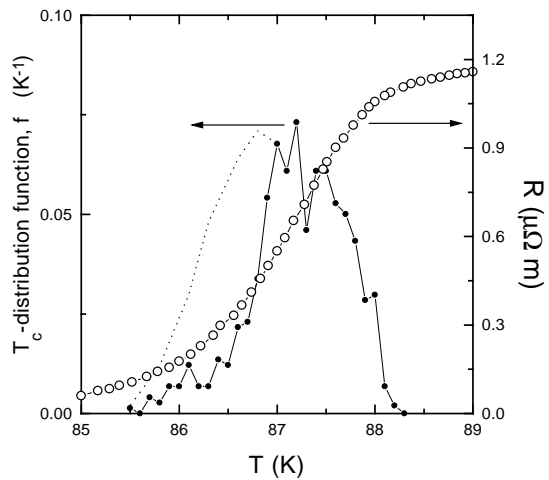,width=8cm}}
\caption{\label{f_distc}
Experimental temperature dependence of resistivity
(circles) for $H=0$ and the
distribution function, $f(T_c)$,
of critical temperature (solid line)
determined from $T_c$-map shown in Fig.~\protect\ref{f_tcmap}.
All the data are for sample~1.
The dotted line shows a plausible shape of the
total distribution function. }
\end{figure}

The method for determination of the
spatial distribution of critical temperature is based
on LTSEM technique\cite{LTSEM,LTSEM1} and is described in detail
in Ref. \onlinecite{ScMic}.
Heating by electron beam elevates the temperature locally
by $\delta T_{\rm{heat}} \le 1$~K causing a change, $\delta\rho$, in
the local resistivity. As a result, a change
in the voltage, EBIV, occurs across the sample biased
by a constant transport current. Temperature dependence
of EBIV has the maximum at some temperature, $T_m$,
corresponding to the maximum in $\delta\rho$.
Thus, the local transition
temperature can be determined as $T_c=T_m+\delta T_{\rm{heat}}/2$.
Scanning the electron beam over the film allows us to
determine the spatial distribution of $T_c$.
In order to remove
the distorting effects associated with thermal diffusion
into adjacent regions of the film a numerical
deconvolution method was used. Figure~\ref{f_ebiv} shows
temperature dependences of EBIV for two adjacent
regions of sample~1 before and after the deconvolution procedure.
After the deconvolution,
both dependences have a pronounced major peak;
its position defines the local $T_c$. It follows from
Fig.~\ref{f_ebiv} that the difference in $T_c$ for two regions
separated by 5~$\mu$m can be as large as 0.7~K.
The method allows the spatial resolution of 2~$\mu$m
and the temperature resolution of 0.2~K.
The \Tc-map for sample~1 is shown in Fig.~\ref{f_tcmap}.
After \Tc-map is determined, one can easily calculate 
distribution function $f(T_c)$ defining the relative
volume occupied by fragments with
given \Tc; $f(T_c)$ for sample~1 is shown in
Fig.~\ref{f_distc}.

Further, using \Tc-map and the expression for conductivity
$\sigma(T,H,T_c)$ of a \Tc-uniform fragment, one can calculate
the spatial distribution of current density in the superconductor.
First, the film is approximated by a square
network of resistors. Then, the set of Kirchoff equations
is solved with respect to 
electric potentials in the nodes of the network.
For this purpose an iterative procedure with
overrelaxation method is used with fixed potentials
of the two opposite sides of
the network.\cite{kir} As 

\noindent
a result, the current density
distribution as well as 
the total resistance
of the superconductor are calculated for any temperature
in the vicinity of the superconducting transition.

As the temperature is lowered, the current density distribution becomes
noticeably inhomogeneous. As a result, some normal-conducting
regions of the film are shunted by surrounding
superconducting regions. $T_c$ of these shunted regions cannot
be measured by the present method. However, one can expect that
ambiguity in their \Tc's would not lead to substantial
errors in results of resistor network calculations. Indeed,
in the high-temperature   part of the superconducting transition,
the conductivity of these regions is known since they
are in the normal state, while at lower temperatures they are
off the main current path and make a minor contribution to
the film resistance. Figure~\ref{f_distc}
represents distribution function $f(T_c)$ determined from the \Tc-map.
The dotted line in the same figure shows a plausible shape of
the total distribution function.

When the spatial distribution of critical temperature
$T_c({\bf r})$ is given, one can estimate the correlation
length of \Tc-inhomogeneity. It is defined from the
correlation function $G(r)$ of \Tc-distribution :
\begin{equation}
G(r)=\frac{\overline{T_c({\bf R}+{\bf r}) T_c({\bf R}) }
- \overline{T_c}^2}
{\overline{T_c^2} - \overline{T_c}^2},
\label{g}
\end{equation}
where averaging is performed over all {\bf R} and all
directions of {\bf r} within the bridge. The value $G=1$
corresponds to full correlation and $G=0$ to the absence
of correlation. For most samples
the correlation function fits very well the exponential
decay, $G(r) \propto e^{-r/r_c}$.

\section{Experimental results}

The parameters of studied YBCO films are presented in
Tab.~1. The fourth column shows the width, $\dtr$, of resistive
transition defined as the doubled dispersion of the Gaussian
fitting $dR/dT$ peak for $H=0$.
The value $\dtr$ equals to approximately 0.8 of the transition width
defined by 10\%-90\% level of normal resistance.

The width $\dtc$ of the experimentally determined
distribution function $f(T_c)$ was calculated by the same
procedure as $\dtr$. For samples marked by ($^*$) the
distribution function had two rather than one peak. In
this case we calculated $\dtc$ as a mean-squared
deviation:
\be
\dtc=2\sqrt{\left\langle (T_c-\bar{T_c})^2\right\rangle }
\label{defdtc}
\ee
where the averaging is performed over the area under
the double-peak Gaussian fitting $f(T_c)$.
Application of Eq.~(\ref{defdtc}) to the distribution
function itself is less reliable because the value of
$\dtc$ is strongly affected by the tails of the
distribution.

Further, we assume that the total broadening of the transition
is caused by summation of homogeneous and inhomogeneous
broadening and the simple relation can be written:
\be
\dtr^2 = \dth^2+\dtc^2,
\label{broad}
\ee
where $\dth$ is the homogeneous broadening of the transition.

The scale $r_c$ of \Tc-inhomogeneity was determined by
fitting the correlation function $G(r)$, Eq.~(\ref{g}),
with an exponential decay, $\exp(-r/r_c)$. Values of
$r_c$ vary much for different samples and depend
primarily on the substrate. This is consistent
with results of x-ray studies which revealed clusters
of dislocations of $\sim$80~$\mu$m size in MgO substrate
used for sample~1. By contrast, sample~4 grown on
NdGaO$_3$ substrate was of higher quality and no large-scale
clusters in the substrate were observed. It should
be noted, that values of $r_c$ in Table~1 can
overestimate the true correlation length of
\Tc-inhomogeneity especially for samples with small $r_c$.
The reason is that $r_c$ is always larger than
the resolution of the experimental method,
$r_{\exp}=2$~$\mu$m.
Presence of \Tc-inhomogeneity  on small scales can be revealed by x-ray
diffraction studies. 

\wide

\vspace{0.0cm}

\noindent
Table 1. Some characteristics of studied YBCO thin film
samples. The transition width, $\dtr$, is defined by the
width of $dR/dT$ peak; $\dtc$ is the dispersion of $T_c$-
distribution; $\dth$ is the intrinsic
broadening of the transition; $\dte$ is the average width
of the local temperature dependence of EBIV; $r_c$ is the
correlation length of $T_c$-distribution;
$\rho_n(T)$ is the linear fit for the temperature
dependence of resistivity in the 150-300~K range.
\vspace{0.5cm}
\noindent
\begin{center}
 \begin{tabular}{|c|c|c|c|c|c|c|c|c|}
 \hline
No.&
Substrate &
$T_c$, K&
$\dtr$,~K &
$\dtc$,~K &
$\dth=$&
$\dte$,~K&
$r_c$, $\mu$m&
$\rho_n$($T$,~K), $\mu\Omega\,m$ \\
&&&&& $\sqrt{\dtr^2-\dtc^2}$ &&&\\
 \hline
 1 & MgO & 86 & 1.5 & 1.2 & 1.1 & 1.1 & 80 & 1.06+0.0035
$T$\\
 \hline
 2 & AlLaO$_3$& 92.8 & 0.4 & 0.3 & 0.3 & 0.3 & - &
0.14+0.003 $T$ \\
 \hline
 3 & AlLaO$_3$& 91.5 & 1.7 &0.8$^*$& 1.5 & 0.9 & 45 & -
\\
 \hline
 4 & \NdGaO & 89 & 1.5 & 0.2 & 1.5 & 0.9 & 6 & 1.21+ 0.02
$T$\\
 \hline
 5 & \NdGaO & 88.5 & 1.9 & 0.3 & 1.9 & 1.7 & 16 &
1.26+0.018 $T$\\
 \hline
 6 & \NdGaO & 87 & 1.7 &0.8$^*$& 1.5 & 0.5 & 33 &
0.96+0.0045 $T$\\
 \hline
 \end{tabular}
~\end{center}
\narrowtext

\noindent
The size of the area where the coherent
scattering of x-ray wave is established has been found to
be 30-100$\AA$ for YBCO films.\cite{JAP,GauzziXray}
This value defines the lower limit for $r_c$.
It is in agreement with the value $r_c \approx 30~\AA$ for the
size of \Tc-uniform fragment in YBCO film deduced from
analysis of experimental data on voltage noise in the superconducting
transition region.\cite{PhC}

The seventh column, $\dte$, shows the average width of the local
temperature dependence of the EBIV which should be
closely related to the homogeneous broadening $\dth$.
Indeed, a good agreement is observed for samples~1, 2
and~5. In the case of samples~3 and 6 a specific shape of
distribution function $f(T_c)$ which invalidates simple
relation~(\ref{broad}) can be responsible for the
deviation. For sample~4 this deviation is probably
related to very short correlation length $r_c$.
The last column in Tab.~1 represents the linear fit for the temperature
dependence of resistivity in the 150-300~K range;
the error in determination of the fit coefficients is 0.2--1\%.

As follows from Tab.~1, inhomogeneous broadening, $\dtc$,
of the resistive transition is of the same order as
homogeneous one, $\dth$. The homogeneous broadening
characterizes the transition width for a fragment of superconducting
film of 2~$\mu$m size. This width can be either an intrinsic
property of a homogeneous superconductor or it can be
associated with \Tc-inhomogeneity  on scales $<2~\mu$m. Large scatter
of $\dth$ in Tab.~1 suggests the presence of small-scale
\Tc-inhomogeneity  at least in the samples with large $\dth$.

Let us now examine the effect of \Tc-inhomogeneity  on the
experimental temperature dependences of conductivity.
Data for samples~1 and~4 with maximal and
minimal $r_c$ will be analyzed.
In order to extract the superconducting contribution,
$\sigma_s(T,H)$, to conductivity
from the measured resistance $R(T,H)$ we
use Eq.~(\ref{sigman}) and data from Tab.~1. The
extracted temperature  dependences of $\sigma_s$ were fitted by two
models: for homogeneous and for \Tc-inhomogeneous
superconductor. For homogeneous superconductor they were
fitted directly by low-field approximation, Eqs.~\reflh,
derived in Sec.~II. The parameters $\sigma_0$, $\kappa$
and \Tc\ were free. For \Tc-inhomogeneous superconductor
the same formulas were used to calculate conductivities
of local fragments with uniform \Tc.
Effective conductivity
$\sigma_s(T,H)$ of the whole sample was calculated by solving resistor
networks based on the measured \Tc-maps. This method has only
two fitting parameters: $\sigma_0$ and $\kappa$.
Theoretical estimate for $\sigma_0$ was obtained in
Sec.~II by comparison of the results of UD model at high
temperatures and Aslamazov-Larkin formula. However, because of
sample imperfections and a large error in determination
of the sample thickness, $\sigma_0$ should be a free
parameter. For studied samples $\sigma_0$ differs from the value
given by Eq.~(\ref{sigma0}) by a factor between 0.6 to
1.5. As follows from formulas of Sec.~II, $\sigma_0$
controls the magnitude of the Cooper pair conductivity, while
$\kappa$ determines the width of the resistive transition.

\begin{figure}[p]
\begin{center}
\psfig{figure=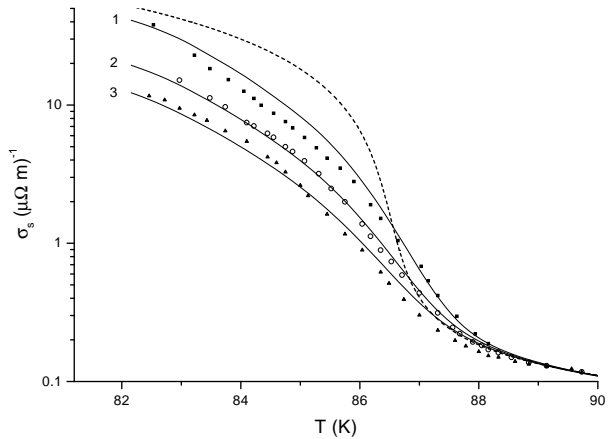,width=8cm}
~\end{center}
\caption{\label{f_3h}
Experimental temperature dependences (symbols) of
superconducting contribution to conductivity, $\sigma_s$,
for sample~1 and their fits (solid lines) calculated
by solving resistor networks based on the measured
spatial distribution of \Tc\ and Eqs.~\protect\reflh
for three magnetic fields:
(1) $H=0.3$~T, (2) $H=0.6$~T, (3) $H=0.9$~T;
the fitting parameter is $\kappa=50$.
The dashed line shows fit for a homogeneous superconductor,
Eqs.~\protect\reflh, for $H=0.6$~T with fitting
parameters $T_c=87.1$~K and
$\kappa=30$. Fits taking account
of \Tc-inhomogeneity  are in a much better agreement
with experiment.}
\end{figure}

The experimental dependences
$\sigma_s(T)$ for sample~1 for three magnetic fields
and their fits by the ``inhomogeneous'' model
are presented in Fig.~\ref{f_3h}.
The dashed line shows the fit by the ``homogeneous'' model
for $H=0.6$~T. 

\begin{figure}[p]
\begin{center}
\psfig{figure=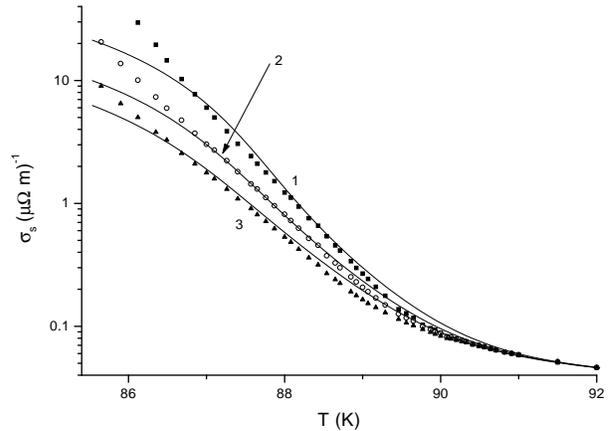,width=8cm}
~\end{center}
\caption{\label{f_sample4}
Experimental temperature dependences (symbols) of
superconducting contribution to conductivity, $\sigma_s$,
for sample~4 and their fits (lines) based on Eqs.~\protect\reflh with
account of \Tc-inhomogeneity  for three magnetic fields: (1) $H=0.3$~T,
(2) $H=0.6$~T, (3) $H=0.9$~T.
Since the correlation length $r_c$ of
\Tc-inhomogeneity  for sample~4 is very small and comparable to the
resolution of \Tc-map, the effective medium approximation
with Gaussian \Tc-distribution is used for calculations.
The fitting parameters are $T_c=90$~K, and $\kappa=30$.}
\end{figure}

\noindent
It can be seen that this model strongly
deviates from the experimental curve. The ``homogeneous'' model
predicts an abrupt rise in conductivity as the temperature decreases
which is not observed on experiment. In ``inhomogeneous''
model this contradiction disappears.

As it can be seen from Tab.~1, the width, $\dtc$, of the measured
\Tc-distribution for sample~4 is substantially
less than the transition width, $\dtr$. We believe that this fact
as well as small $r_c$ are related to
presence of \Tc-inhomogeneities on scales less
than the experimental resolution, $r_{\exp}$.
In this case calculations based on the measured
\Tc-map are not reliable. Instead,
in order to calculate the effective conductivity for sample~4, we
used EMA and a Gaussian \Tc-distribution function. The
results for three magnetic fields are shown in
Fig.~\ref{f_sample4}. Additional fitting parameters, the
average and the dispersion of Gaussian distribution, were
found to be $T_{c0}=89.1$~K, and $\dsp=1.3$~K.

Presence of small-scale \Tc-inhomogeneities
is probably the reason for difference
in $\kappa$ determined from fitting the experimental $\sigma_s(T)$
dependences for different samples.
For sample~4 (Fig.~\ref{f_sample4}) the best fits based on the EMA are
obtained with $\kappa=30$,
while for sample~1 (Fig.~\ref{f_3h}) the best fits
based on the measured \Tc-map give $\kappa=50$.
The high $\kappa$ in the latter case leads
to additional broadening of the transition
compensating lack of information about
small-scale \Tc-inhomogeneities.
Thus, the value $\kappa=30$ is more reliable
and it is used for calculations presented in Sec.~III.

To summarize, there are two ways to take \Tc-inhomogeneity into
account: direct resistor network calculations based on
\Tc-map, and EMA along with a Gaussian \Tc-distribution. The
resistor network calculations have the advantage of using
actual spatial distribution of \Tc\ in the sample. It has
the information about location of regions with various
\Tc\ allowing the calculation of percolative current
distribution in given HTSC film. On the other hand, the
drawback of this model is that the \Tc-map is
measured with finite spatial resolution.
Thus, one should use either \Tc-map
or EMA for large and small values of the correlation
length, $r_c$, respectively.

In Fig.~\ref{f_sample4} all models
significantly deviate from the experimental data at
sufficiently low temperatures. We explain this deviation by the vortex
pinning which comes into play for low temperatures and prevents
the dissipation associated with flux flow. UD model
does not take the pinning into account and, thus, overestimates
the dissipation rate. We believe that in the
low--temperature part of the superconducting transition it is the
strength and concentration of pinning centers
rather than \Tc-distribution that
controls the transport properties.

It is well-known that while the resistance of an
inhomogeneous system is determined by the second moment
of current distribution, the resistance fluctuations are
determined by the fourth moment. Therefore the resistance
noise is far more sensitive to the presence of all kind
of inhomogeneities than the resistance itself.\cite{Bergman}
This means that although this
work presents analysis of the transport properties only,
one can expect far stronger effect of \Tc-inhomogeneity  on the noise
properties of superconductors. Even simple analysis not
involving any particular dependence of local conductivity on $T$
and $H$ shows a strong effect of \Tc-inhomogeneity  on the level of
thermodynamic noise \cite{TDN} and the noise associated
with fluctuations of local \Tc.\cite{PhC}

The properties of single crystals differ much from those
of thin films and need a special consideration. It is
generally believed that small transition width, $0.1-0.3$~K,
in zero magnetic field observed in single crystals proves
their high homogeneity. Therefore, the experimental data
on single crystals are often used to get insight into
fundamental intrinsic properties of superconductors. Nevertheless,
recent theoretical and experimental investigations make
their homogeneity, in particular, \Tc-homogeneity
questionable. It is predicted\cite{Gurevich} that various
extended structural defects, e.g., dislocations can give
rise to formation of the extended regions with enhanced
\Tc\ nearby.
Studies of the influence of oxygen stoichiometry on the
magnetization curves of YBCO crystals suggest that
the so called peak effect
widely observed in HTSC crystals is associated with
the presence of local regions with reduced oxygen content,
and, hence, reduced \Tc.\cite{Kupfer,Zhukov}
Presence of non-uniform \Tc-distribution in 
\YBCO\ and Bi$_2$Sr$_2$CaCu$_2$O$_8$ crystals 
follows from experimental data on in-plane 
magnetoresistivity anomalities.\cite{MosSST} 
Further, large-scale spatial variations of
oxygen composition, implying variations of \Tc, were
observed\cite{Browning} in YBCO single crystals by x-ray studies.
However, the spatial
scale of \Tc-inhomogeneities in crystals often has a value
comparable to the size of the sample.\cite{Browning,MosSST}
In such a case, despite
a wide distribution of \Tc\ over the sample, the
superconducting transition can be very sharp because of a
percolation over high-\Tc\ regions along one of the
sample edges. Unfortunately,
such situations cannot be properly treated in the frame of
the effective medium approach because it assumes purely
uncorrelated \Tc-distribution. EMA can neither be
applicable to describe wires of higher \Tc\ near extended
structural defects.\cite{Gurevich} Thus, we do not expect that
the results of this work would be applicable to HTSC
crystals. Nevertheless, there are grounds to believe that
inhomogeneity of crystals strongly manifests itself in
their properties and deserves a detailed analysis.

\section{Conclusions}

\Tc-inhomogeneity of YBCO films is directly
demonstrated by measuring spatial distributions of \Tc\ by
low-temperature SEM with 2~$\mu$m resolution. The
dispersion of \Tc-distribution was found to be of the order
of 1~K which is comparable to the resistive transition width. This
result indicates inhomogeneous broadening of the
resistive transition for the films studied.

We obtain a non-explicit expression for Cooper pair conductivity
$\sigma_s(T,H,T_c)$
of a homogeneous superconductor, which is valid throughout
the transition region for magnetic fields $H \ll H_{c2}(0)$.
For \YBCO, it can be reduced to an explicit expression
for fields $H \ll 0.1 H_{c2}(0)$.

We find that the error in the apparent value of $\sigma_s(T,H,T_c)$
due to \Tc-inhomogeneity
is maximal for low magnetic fields and temperatures close to \Tc.
For YBCO films with a Gaussian \Tc-distribution with 1~K dispersion,
ignoring \Tc-inhomogeneity  leads to more than 30\%
error in $\sigma_s$
in the region restricted to temperatures $|T-T_c|<0.5$~K
and magnetic fields $H < 1$~T. Thus, it is necessary to be
cautious when carrying out quantitative analysis of experimental
data in the transition region. One of the following
is recommended: (i)~carry out all measurements
beyond the region affected by \Tc-inhomogeneity, i.e., at
very high magnetic fields or at temperatures far from
\Tc; (ii) take \Tc-inhomogeneity into account by
measuring \Tc-spatial distribution or, at least, by assuming
a Gaussian distribution and using EMA or similar
approximation.

Finally, it should be noted that the boundaries of
$H$--$T$ plane region
affected by \Tc-inhomogeneity  are determined not only
by microscopic superconducting parameters, but also by
material parameters such as dispersion and correlation
length of \Tc-inhomogeneity. Nevertheless, a transition
width of the order of 1~K seems typical for YBCO
films, while Bi-based films usually have even broader
transition. Thus, the presented results are likely to
be relevant to most HTSC films.

\acknowledgements

The work is supported by Russian Program on
Superconductivity, Projects 98031 and~96071. The authors wish to
thank V.~A.~Solov'ev, Yu.~M.~Galperin, V.~I.~Kozub, and A.~I.~Morosov, for
helpful discussions, S.~F.~Karmanennko for sample
fabrication, and J.~Alexander for help in preparation of the
manuscript.

\widetext

\begin{thebibliography}{50}

\bibitem[*]{0} E-mail address: shantsev@theory.ioffe.rssi.ru

\bibitem{PhysCSiddique} Rezaul K. Siddique, Physica C
{\bf 228}, 365 (1994).

\bibitem{rusGasumants} V.~E.~Gasumants, S.~A.~Kazmin,
V.~I.~Kaidanov, V.~I.~Smirnov,
Yu.~M.~Baikov, and Yu.~P.~Stepanov, Sverhprovod. Fiz. Him.
Teh. {\bf 4}, 1280 (1991).

\bibitem{Browning} V.M. Browning, E.F. Skelton, M.S. Osofsky,
S.B. Qadri, J.Z. Hu, L.W. Finger, and P. Caubet,
Phys. Rev. B {\bf 56}, 2860 (1997).

\bibitem{cations} N.~A.~Bert, A.~V.~Lunev,
Yu.~G.~Musikhin, R.~A.~Suris, V.~V.~Tret'yakov,
A.~V.~Bobyl, S.~F.~Karmanenko, and A.~I.~Dedoboretz, Physica
C {\bf 280}, 121 (1997).

\bibitem{3methods} M. E. Gaevski, A. V. Bobyl, D. V.
Shantsev, Y. M. Galperin, V. V.
Tret'yakov, T. H. Johansen, and R. A. Suris,
J. Appl. Phys. {\bf 84}, 5089 (1998).

\bibitem{pressure} C.~C.~Almasan, S.~H.~Han, B.~W.~Lee,
L.~M.~Paulius, M.~B.~Maple, B.~W.~Veal, J.~W.~Downey, A.~P.~Paulikas,
Z.~Fisk, and J.~E.~Schriber, Phys. Rev. Lett. {\bf 69}, 680 (1992).

\bibitem{Gurevich} A.Gurevich and E.A.Pashitskii, Phys.
Rev. B  56, 6213 (1997).

\bibitem{JAP} A. V. Bobyl, M.E.Gaevski, S.F.Karmanenko,
R.N. Kutt, R.A.Suris, I.A.~Khrebtov, A.~D.~Tkachenko, and A.~I.~Morosov,
J. Appl. Phys. {\bf 82}, 1274 (1997).

\bibitem{pinningWen} H.~H.~Wen, Z.~X.~Zhao, Y.~G.~Xiao,
B.~Yin, J.~W.~Li,
Physica C, {\bf 251}, 371 (1995).

\bibitem{Kupfer}
H. Kupfer, Th. Wolf, C. Lessing, A. A. Zhukov,
X. Lancon, R. Meier-Hirmer, W. Schauer, and H. Wuhl,
Phys. Rev. B {\bf 58}, 2886 (1998).

\bibitem{Zhukov}
A. A. Zhukov, H. Kupfer, G. Perkins, L. F. Cohen, and A. D. Caplin,
S. A. Klestov, H. Claus, V. I. Voronkova, T. Wolf and H. Wuhl,
Phys. Rev. B {\bf 51}, 12704 (1995).

\bibitem{Osofsky} M. S. Osofsky, J. L. Cohn,
E.~F.~Skelton, M.~M.~Miller, R.~J.~Soulen, Jr., and S.~A.~Wolf,
T.~A.~Vanderah, Phys. Rev. B {\bf 45}, 4916 (1992).

\bibitem{PomVidal} A. Pomar, M.~V.~Ramallo, J. Mosqueira,
C.~Torron, and F.~Vidal, Phys. Rev. B {\bf 54}, 7470 (1996);
J~ Mosqueira, A.~Pomar, A.~Diaz, J.~A.~Veira, and
F.~Vidal, Physica C {\bf 225}, 34 (1994).

\bibitem{Lang} W.~Lang, Physica C {\bf 226}, 267 (1994).

\bibitem{Lang95} W.~Lang, G.~Heine, W. Kula, and
R.~Sobolewski, Phys. Rev. B {\bf 51}, 9180 (1995).

\bibitem{MosSST}  J. Mosqueira, S. R. Curras, C. Carballeira, 
M.~V.~Ramallo, Th.~Siebold, C.~Torron, J.~A.~Campa,
I.~Rasines, and F.~Vidal, Supercond. Sci. Technol. {\bf 11},
821 (1998).

\bibitem{UD} S.~Ullah and A.~T.~Dorsey, Phys. Rev. B {\bf 44},
262 (1991).

\bibitem{Welp} U.~Welp, S.~Flesher, W.K.~Kwok,
R.A.~Klemm, V.M.~Vinokur, J.~Downey, B.~Veal, and
G.W.~Crabtree, Phys. Rev. Lett. {\bf 67}, 3180 (1991).

\bibitem{LD} W.E.Lawrence and S.Doniach, in: Proc. 12th
Int. Conf. on Low Temperature Physics, Kyoto, 1970, ed. E.~Kanda
(Keigaku, Tokyo, 1971), p.361.

\bibitem{UDerror} There is a misprint in the original
paper of \UD\cite{UD} in the definition of $A_n$ following
formula~(4.1): the exponent after the square
brackets should be -1/2 instead of 1/2.
The existence of this misprint is clear to see from the simple
fact that each term in the perturbation series must be smaller
than the preceding one. There is also a misprint in
the next formula~(4.2): the second denominator in square brackets
$\tilde \epsilon_H(1+d^2 \tilde \epsilon_H)$
should be put under the square root. For discussion
of misprints in Ref.~\onlinecite{UD} see also
``Note added to the proof'' in Ref.~\onlinecite{rice}.

\bibitem{EM} In the first order the
Euler-Maklaurin formula reads $\sum_{n=a}^b f(n) = \int_a^b f(n)\ dn+ 
f(a)/2 + f(b)/2$.

\bibitem{bilayer} M. V. Ramalio, A. Pomar and
F\'elix Vidal, Phys. Rev. B {\bf 54}, 4341 (1996).

\bibitem{rice} J. P. Rice, J .Giapintzakis, D. M. Ginzberg,
and J. M. Mochel, Phys. Rev B {\bf 44}, 10158 (1991).

\bibitem{xi}
W.~Lang, G.~Heine, P.~Schwab, X.~Z.~Wang, and D.~B\"auerle,
Phys. Rev. B {\bf 49}, 4209 (1994);
W.~Holm, \"O.~Rapp, C.~N.~Johnson, and U.~Helmersson,
Phys. Rev. B {\bf 52}, 3748 (1995);
A.~Pomar, A.~Diaz, M.~V.~Ramalio, C.~Torron, J.~A.~Veira,
 and F\'elix Vidal, Physica C {\bf 218}, 257 (1993).

\bibitem{AL} L. G.~Aslamazov and A. I.~Larkin, Phys. Lett.
{\bf 26A}, 238 (1968).


\bibitem{MT} K.~Maki, Prog. Theor. Phys. {\bf 39}, 897 (1968);
R.S.~Thompson, Phys. Rev. B {\bf 1}, 327 (1970).

\bibitem{MTargue}
K. Semba and A. Matsuda, Phys. Rev. B {\bf 55}, 11103 (1997).

\bibitem{Dyhne} A. M. Dyhne, Zh. Eksper i Teor. Fiz. {\bf
59}, 110 (1970).

\bibitem{EMA} R.~Landauer, J. Appl. Phys. 23, 779 (1956).

\bibitem{kir}S. Kirckpatrick, Rev. Mod. Phys. {\bf 45},
574 (1973).

\bibitem{meltingPRB} C. M. Aegerter, S. T. Johnson, W. J. Nuttall,
S. H. Lloyd, M. T. Wylie, M. P. Nutley, E. M. Forgan, R. Cubitt,
S. L. Lee, D. McK. Paul, M. Yethiraj, H. A. Mook, Phys. Rev. B
{\bf 57}, 14511 (1998).

\bibitem{meltingPRL} U. Welp, J. A. Fendrich, W. K. Kwok, G.
W. Crabtree, B. W. Veal, Phys. Rev. Lett. {\bf 76}, 4809 (1996).


\bibitem{GauzziXray} A. Gauzzi and D. Pavuna, Appl. Phys.
Lett. {\bf 66} 1836 (1995).

\bibitem{PhC} A.~V.~Bobyl, M.~E.~Gaevski, I.~A.~Khrebtov,
S.~G.~Konnikov, D.~V.~Shantsev,
V.~A.~Solov'ev, R.~A.~Suris, and A.~D.~Tkachenko, Physica C
{\bf 247}, 7 (1995).

\bibitem{LTSEM} R. P. Huebener, in: {\em Advances in
Electronics and Electron
Physics}, {\bf 70}, ed. by P.~W.~Hawkes (Academic, New
York, 1988), p.1.

\bibitem{LTSEM1}R.~Gross and D.~Koelle, Rep. Prog. Phys. {\bf 57}, 651
(1994).

\bibitem{ScMic} M. E. Gaevski, A. V. Bobyl, S.~G.~Konnikov,
D.~V.~Shantsev, V.~A.~Solov'ev, and R.~A.~Suris, Scanning Microscopy, {\bf 10},
679 (1996).

\bibitem{Bergman} D. J. Bergman, Phys. Rev. B {\bf 39},
4598 (1989).

\bibitem{TDN} N. V. Fomin and D. V. Shantsev, Tech. Phys.
Lett. {\bf 20}, 50 (1994) (Pis'ma Zh. Tekh. Fiz. {\bf 20}, 9 (1994)).

\end{thebibliography}
\end{document}